\documentclass[10pt,twocolumn,letterpaper]{article}

\usepackage{cvpr}              
\usepackage{colortbl}
\usepackage{bm}
\usepackage{footmisc}
\usepackage{multirow}
\usepackage[ruled]{algorithm2e} 

\definecolor{color_1st}{rgb}{0.82, 0.71, 0.58}
\definecolor{color_2nd}{rgb}{0.945, 0.87, 0.74}
\definecolor{color_3rd}{rgb}{1, 0.98, 0.84}

\definecolor{cvprblue}{rgb}{0.21,0.49,0.74}
\usepackage[pagebackref,breaklinks,colorlinks,allcolors=cvprblue]{hyperref}
\def\paperID{*****} 
\def\confName{CVPR}
\def\confYear{2025}

\title{CityGo: Lightweight Urban Modeling and Rendering with Proxy Buildings and Residual Gaussians}

\author{
Weihang Liu$^{1,4,*}$, Yuhui Zhong$^{2,*}$, Yuke Li$^{1,*}$, Xi Chen$^{1}$, Jiadi Cui$^{1,5}$, Honglong Zhang$^{3}$, \\
Lan Xu$^{1}$, Xin Lou$^{1,4}$, Yujiao Shi$^{1}$, Jingyi Yu$^{1}$, Yingliang Zhang$^{2}$\\
\text{$^{1}$ShanghaiTech University},
\text{$^{2}$DGene},
\text{$^{3}$Migu Cultural Technology Co.,Ltd}, \\
\text{$^{4}$GGU Technology Co., Ltd},
\text{$^{5}$Stereye}
}

\begin{document}
\twocolumn[{%
\renewcommand\twocolumn[1][]{#1}%
\maketitle
\includegraphics[width=1.0\linewidth]{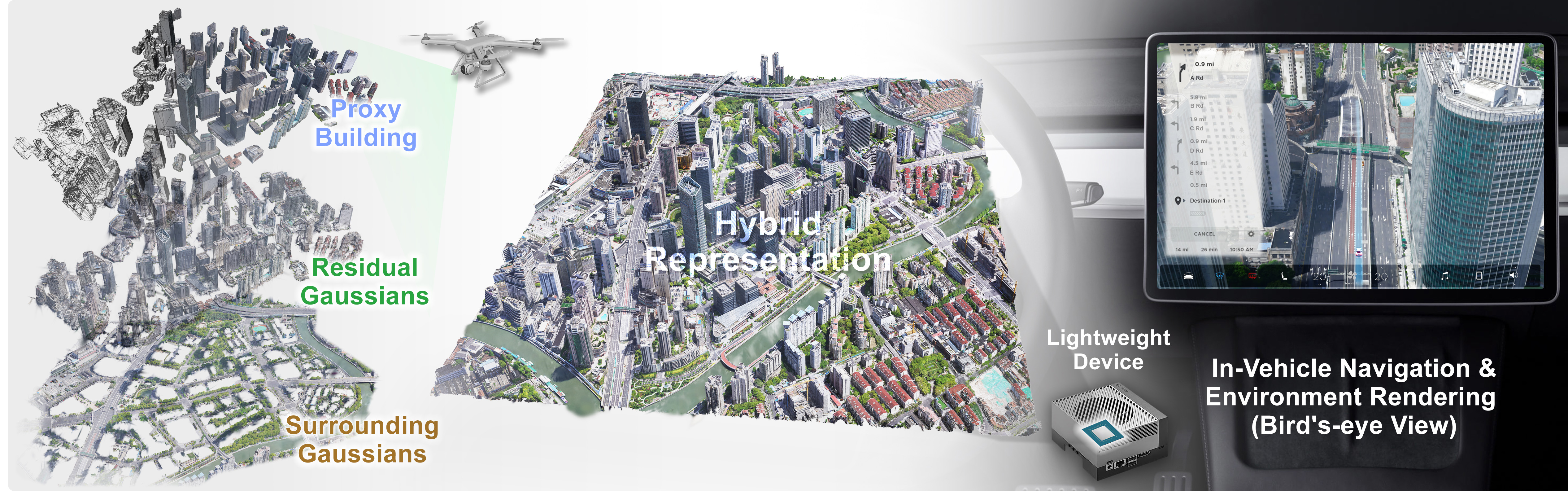}
    \captionof{figure}{We present CityGo, an explicit and efficient framework for high-fidelity rendering of large-scale urban scenes. 
    By combining proxy buildings, residual Gaussians, and surrounding Gaussians, we enable efficient, high-quality urban scene rendering on lightweight devices for applications such as in-vehicle navigation and aerial perception.}
\label{fig:teaser}
}]
\def\thefootnote{*}\footnotetext{Equal Contribution.}
\begin{abstract}
Accurate and efficient modeling of large-scale urban scenes is critical for applications such as AR navigation, UAV-based inspection, and smart city digital twins. While aerial imagery offers broad coverage and complements limitations of ground-based data, reconstructing city-scale environments from such views remains challenging due to occlusions, incomplete geometry, and high memory demands. Recent advances like 3D Gaussian Splatting (3DGS) improve scalability and visual quality but remain limited by dense primitive usage, long training times, and poor suitability for edge devices.
We propose CityGo, a hybrid framework that combines textured proxy geometry with residual and surrounding 3D Gaussians for lightweight, photorealistic rendering of urban scenes from aerial perspectives. Our approach first extracts compact building proxy meshes from MVS point clouds, then uses zero-order SH Gaussians to generate occlusion-free textures via image-based rendering and back-projection. To capture high-frequency details, we introduce residual Gaussians placed based on proxy-photo discrepancies and guided by depth priors. Broader urban context is represented by surrounding Gaussians, with importance-aware downsampling applied to non-critical regions to reduce redundancy.
A tailored optimization strategy jointly refines proxy textures and Gaussian parameters, enabling real-time rendering of complex urban scenes on mobile GPUs with significantly reduced training and memory requirements. 
Extensive experiments on real-world aerial datasets demonstrate that our hybrid representation significantly reduces training time, achieving on average $1.4\times$ speedup, while delivering comparable visual fidelity to pure 3D Gaussian Splatting approaches. Furthermore, CityGo enables real-time rendering of large-scale urban scenes on mobile consumer GPUs, with substantially reduced memory usage and energy consumption. 
\textit{Project page: 
\href{https://citygo-weihang.github.io/page/}{\url{https://citygo-weihang.github.io/page/}}
}

\end{abstract}    
\section{Introduction}
\label{sec:intro}

Large-scale urban modeling and rendering are foundational technologies for a wide range of applications, including urban planning, autonomous navigation, and the creation of digital twins for smart cities. Among various scene elements, accurate modeling of buildings is particularly critical, as they define much of a city's structural layout and visual identity.

While ground-based imagery has been widely used in urban modeling, it often faces challenges when dealing with tall structures and complex occlusions. Aerial imagery, such as that captured by unmanned aerial vehicles (UAVs), offers broader scene coverage, capturing rooftops and spatial layouts inaccessible from the ground (Fig.~\ref{fig:Gallery1}, Fig.~\ref{fig:Gallery2}). As UAV data becomes increasingly accessible, it provides a promising avenue for scalable and efficient urban modeling.
At the same time, there is a growing demand for real-time and lightweight rendering of city-scale models on edge devices like UAVs, smartphones, and AR glasses. 
These use cases impose stringent constraints on memory, power, and latency, calling for compact yet photorealistic scene representations.

Traditional geometry-based approaches, such as Structure-from-Motion (SfM)~\cite{SfM1, SfM2} and Multi-View Stereo (MVS)~\cite{MVS1, MVS2}, have been the backbone of city-scale reconstruction from aerial imagery. However, they often produce fragmented point clouds and noisy meshes, particularly in textureless areas or under lighting variations, due to local feature matching failures. The complexity of urban geometry, reflective surfaces, and vegetation further exacerbates these limitations.

To overcome these issues, learning-based methods like Neural Radiance Fields (NeRF)~\cite{NeRF} leverage volumetric representations to generate high-quality renderings. However, NeRF models require long training times and offer limited runtime performance. Their implicit, volumetric nature also makes them difficult to compress, edit, or deploy on resource-constrained devices.

More recently, 3D Gaussian Splatting (3DGS)~\cite{3DGS} has emerged as an efficient alternative, representing scenes with explicit Gaussian primitives. 3DGS supports differentiable rendering with view-dependent radiance and has been extended to large-scale settings using divide-and-conquer~\cite{CityGS, CityGS_v2, DoGaussian, VastGaussian} and Level-of-Detail (LoD) \cite{letsgo, HierarchicalGS} strategies. However, modeling an entire city with 3DGS can demand hundreds of millions of Gaussians and consume tens of gigabytes of GPU memory, far beyond the capacity of mobile platforms. Moreover, capturing high-frequency textures often requires dense Gaussian layering, which introduces visual redundancy and blurring in regions with simpler geometry.

In this paper, we propose CityGo, a hybrid and lightweight modeling framework for photorealistic and efficient rendering of large-scale urban scenes from aerial perspectives. Our approach combines textured proxy geometry with residual and surrounding 3D Gaussians, achieving a practical balance between geometric fidelity, texture sharpness, and computational efficiency.

We begin by generating clean, compact proxy building meshes from MVS point clouds. To enable fast appearance initialization, we employ 3D Gaussian Splatting with zero-order spherical harmonics (SH). The resulting Gaussians are segmented on a per-building basis and used to generate occlusion-free textures through image-space rendering and back-projection, avoiding the directional ambiguities inherent in traditional texture mapping. Gaussians that lie outside building regions are retained as surrounding Gaussians, which capture the broader urban context—including roads, vegetation, and other environmental elements—for improved scene realism.

To enhance visual fidelity without incurring excessive memory costs, we introduce residual Gaussians, which are selectively placed in regions exhibiting noticeable color discrepancies between the proxy-rendered images and the original input photos. These residuals preserve high-frequency appearance details while avoiding redundancy with the mesh. Their placement is further guided by depth maps inferred from the proxy geometry to maintain geometric consistency. For less salient regions, such as distant foliage, roads, or background structures, we apply importance-aware Gaussian downsampling to the surrounding Gaussians, reducing computational overhead while preserving perceptual quality.

Finally, we propose a tailored optimization strategy that jointly refines proxy textures via differentiable rendering and optimizes the positions, opacities, and densities of both residual and surrounding Gaussians. This hybrid representation enables real-time rendering of city-scale scenes even on mobile consumer GPUs, while maintaining high visual quality and structural coherence.

Our main contributions are summarized as follows:

\begin{itemize}
    \item We propose CityGo, a hybrid representation that combines textured proxy buildings with residual and surrounding Gaussians for accurate, scalable aerial urban modeling.

    \item By initializing with zero-order SH Gaussians and employing a TwinTex strategy for occlusion-free texture generation, CityGo achieves significantly reduced training time compared to traditional 3DGS-based pipelines.
    
    \item CityGo supports real-time rendering on mobile GPUs with low memory and energy footprints, enabling practical deployment in AR navigation, UAV inspection, and digital twin applications.
\end{itemize}

\section{Related Work}
\label{sec:related_work}

\paragraph{Conventional Explicit 3D Scene Reconstruction.}
Traditional pipelines like Structure-from-Motion (SfM)\cite{SfM1, SfM2}, Multi-View Stereo (MVS)\cite{MVS1, MVS2}, and texture mapping have been widely adopted for 3D urban reconstruction. However, they struggle in complex environments due to reliance on multi-view correspondences, often resulting in noisy, incomplete point clouds—especially around reflective surfaces, repetitive textures, and intricate geometries. The generated dense meshes are memory-intensive, and texture mapping suffers from artifacts such as blurring and ghosting, as it depends on accurate geometry. To reduce complexity for edge devices, mesh simplification methods, such as quadric error metrics~\cite{19}, planar proxies~\cite{20,21,22,23,24}, and primitive extraction~\cite{25,26,27}, are commonly used. Yet these techniques assume clean inputs and often amplify distortions when applied to imperfect reconstructions. Balancing geometric accuracy, visual fidelity, and efficiency thus remains a key challenge for explicit methods.

\begin{figure*}
    \centering
    \includegraphics[width=1.0\linewidth]{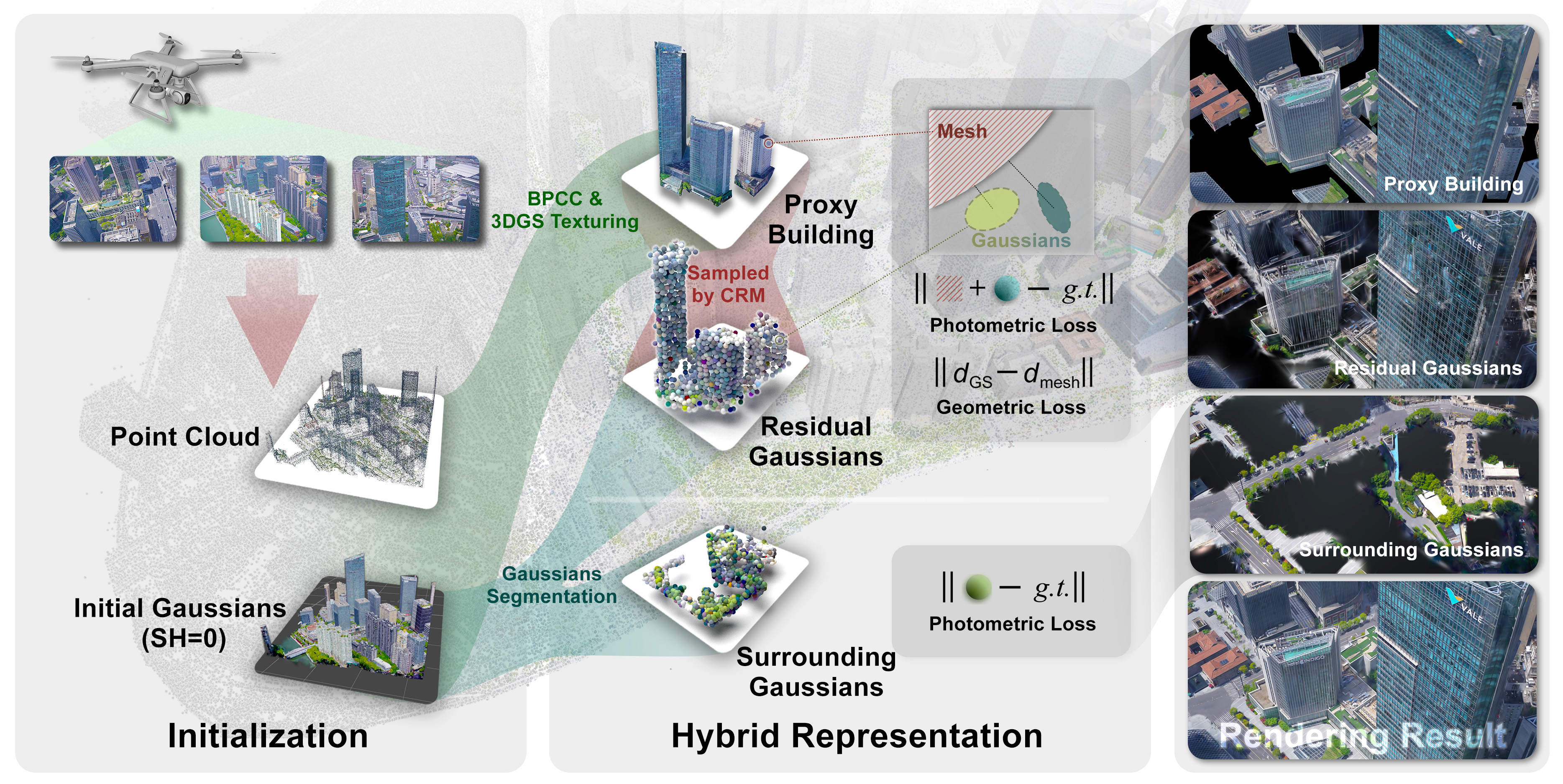}
    \caption{
    Overview of our hybrid representation for large-scale urban scenes. We begin by generating dense point clouds from aerial images and initializing zero-order SH Gaussians to capture the entire scene. Buildings and surrounding areas are then segmented and processed separately. For buildings, we adopt a hybrid representation of textured proxy meshes and residual Gaussians (Sec.\ref{sec:3}), while simplified Gaussians are used for the surroundings (Sec.\ref{sec:4}). The final model enables photorealistic rendering with significant speedups for cinematic or real-time performance on lightweight devices.
    }
    \label{fig:pipeline}
    \vspace{-10pt}
\end{figure*}

\paragraph{Large-Scale Neural Scene Representations.}
Neural Radiance Fields (NeRF)\cite{NeRF} marked a major step in neural rendering, with large-scale variants like Block-NeRF\cite{block_nerf}, ScaNeRF~\cite{ScaNeRF}, and Mega-NeRF~\cite{Mega-NeRF} employing divide-and-conquer strategies for photorealistic reconstruction. However, their high training and rendering costs limit practical deployment. 3D Gaussian Splatting (3DGS)\cite{3DGS} offers a more efficient alternative by explicitly modeling scenes with anisotropic Gaussians, enabling faster optimization and rendering. Subsequent works have improved 3DGS in terms of anti-aliasing\cite{mip_splatting}, memory usage~\cite{mini_splatting}, and adaptive rendering~\cite{scaffold, OctGS}. As a result, recent large-scale systems~\cite{DoGaussian, GrendelGS, FlashGS, CityGS, CityGS_v2} increasingly adopt 3DGS. For instance, VastGaussian~\cite{VastGaussian} improves partitioning and appearance modeling, Hierarchical 3DGS~\cite{HierarchicalGS} uses LoD for street-scale rendering, and LetsGo~\cite{letsgo} incorporates LiDAR for garage-scale scenes. Nonetheless, real-time rendering of city-scale scenes on edge devices remains challenging due to the large number of Gaussians and the resulting memory footprint.

\paragraph{Hybrid Scene Representations.}
Hybrid 3D representations that blend meshes, point clouds, NeRFs, and 3D Gaussian Splatting (3DGS) have shown promise for modeling complex scenes. For instance, PointNeRF~\cite{PointNeRF, PointNeRF++} fuses sparse MVS point clouds with implicit fields, while methods like Plenoxels~\cite{Plenoxel}, TensoRF~\cite{TensoRF}, and Instant-NGP~\cite{NGP} enhance efficiency using explicit volumetric grids. More recently, hybrid mesh-Gaussian approaches \cite{meshgs, MeshGSRoom, dreammesh4d} have achieved photorealistic results in object- or human-scale settings, with HERA~\cite{HybridAvatar} and SplattingAvatar~\cite{SplattingAvatar} demonstrating realistic avatar rendering via mesh-guided splatting. These works typically rely on high-quality 3D templates, such as head~\cite{3DMM2, DECA} or body~\cite{FULLbody1} models, which are unavailable for urban-scale environments. To address this, we propose a hybrid representation tailored for large-scale city modeling: structured proxy meshes for buildings, residual Gaussians for fine texture refinement, and sparse Gaussians for surrounding context. This design enables visually coherent and efficient rendering across both desktop and edge hardware.

\section{Hybrid Representation for 3D Buildings}
\label{sec:3}

Given multi-view images captured by UAV drones and building segmentation data derived from geographic information system (GIS), our method, CityGo, models urban scenes using a hybrid representation that separates structured and unstructured components. 
For structured elements such as buildings, we introduce a novel representation combining textured proxy meshes with 3D Gaussians,
enabling photorealistic rendering with high efficiency. Unstructured surroundings are modeled using sparse 3D Gaussians that capture complex, ambient visual details. 
An overview of our pipeline is shown in Fig.~\ref{fig:pipeline}, with each component detailed below.

\subsection{Building Point Cloud Completion}
\label{sec:BPCC}
We start by estimating camera parameters and generating dense point clouds of urban scenes using off-the-shelf Structure-from-Motion and Multi-View Stereo methods~\cite{AgisoftPhotoScan2016}. 
These point clouds are used to initialize a 3D Gaussian Splatting (3DGS) model. 
This process yields both dense point clouds and 3D Gaussians that represent the entire scene. 
Using the GIS-derived building masks, we segment both the dense point clouds and Gaussians to isolate individual buildings from their surroundings. 
However, these building-specific point clouds are often incomplete due to occlusions, repetitive textures, and surface reflections. These challenges hinder downstream mesh extraction and texturing. To address this, we propose a Building Point Cloud Completion (BPCC) method that reconstructs complete, hole-free point clouds from dense MVS data.

Inspired by \cite{PROXY}, we first estimate a layer-based proxy geometry from an input point cloud, and then sample points from the surface of this proxy geometry to fill in missing regions. 
The method \cite{PROXY} proceeds by slicing the sparse point cloud uniformly along the vertical axis into a sequence of horizontal layers
$\mathcal{L} = \{L_1, \cdots, L_n\}$, with $L_1$ denoting the topmost layer and $L_n$ the bottom. For each layer $L_i$, a global projected point set $P_i$ is constructed by projecting all 3D points from layer $L_i$ and above onto the plane of $L_i$. A structural contour $C_i$ for layer $L_i$ is then derived by computing the convex hull of $P_i$. 
To construct the proxy geometry, a subset of these contours is selected from top to bottom to form a set of dominant structural contours $\mathcal{S} = \{S_j \mid j = 1, \dots, m\}$, where $m \leq n$.
For each pair of adjacent contours $(S_j, S_{j+1})$, a volumetric segment is generated by extruding the region enclosed by $S_j$ downward to the level of $S_{j+1}$. The union of these volumetric segments forms the final layered proxy geometry. For further implementation details, we refer readers to \cite{PROXY}.

However, directly adopting the method from \cite{PROXY} poses several challenges. 
Point clouds generated via Structure-from-Motion (SfM) are often sparse and noisy, leading to inaccurate contour estimation. Additionally, convex hull-based representations struggle to capture complex architectural structures, such as concave building layouts or adjacent high-rises (e.g., twin towers), which are common in urban buildings. To overcome these limitations, we propose three key improvements: (1) using a dense point cloud as input, (2) applying clustering to partition the projected 2D points within each layer, and (3) leveraging the alpha shape algorithm~\cite{alphashape} for more flexible and accurate contour extraction.
While our method also vertically slices the dense point cloud into layers, it differs in both direction and processing strategy. Specifically, we compute contours starting from the bottom layer $L_n$ and proceed upward. Rather than precomputing all contours and then selecting a subset, we directly cluster the projected 2D points within each layer, compute contours for each cluster, and identify dominant contours. This bottom-up, cluster-aware approach yields a more faithful reconstruction of layered proxy geometry in urban scenes.

First, we apply the DBSCAN algorithm~\cite{DBSCAN} to cluster the global projected point set $P_n$ on the bottommost layer, yielding subsets $\{P_n^k \mid k = 1, \cdots, o\}$. For each $P_n^k$, we compute its contour $C_n^k$ using the alpha-shape algorithm. These contours form the initial set of dominant structural contours $\mathcal{S}$, marked as current dominant contour $\hat{S}_n^{k}$. As we project dense points to each layer, the set $P$ on each layer satisfies $P_i \subset P_{i+1}$. Consequently, for each upper layer $i = 1, \cdots, n-1$, the point set $P_i$ inherits the clustering results from the lower adjacent layer, defined as $\hat{P}_i^{k} = P_i \cap P_{i+1}^k$, and is subjected to re-clustering. A new dominant contour is introduced whenever either of two specific conditions indicating significant vertical structural changes is met: (1) $\hat{P}_i^{k}$ splits into multiple clusters, or (2) $\hat{P}_i^{k}$ forms a single cluster but its alpha-shape contour $\hat{C}_i^{k}$ significantly reduces in area compared to the previously identified dominant contour, specifically $\frac{Area(\hat{C}_i^{k})}{Area(\hat{S}_{i+1}^{k})} \leq \gamma$, where $\gamma$ is a fixed threshold. When these conditions are satisfied, new alpha-shape contours are computed and added to $\mathcal{S}$, updating the current dominant contours. If neither condition is met, the contour from the lower layer is retained.
This bottom-up process repeats for all layers, resulting in the final set of dominant contours $\mathcal{S}$. Using these contours, we estimate the layered proxy geometries as in \cite{PROXY}. Finally, we sample points from the derived proxy geometries to fill in gaps on the building point cloud’s bottom and side surfaces. The detailed procedure for the proposed approach, named BPCC, is presented in \textit{Appendix}.\ref{app:bccp}.

\begin{figure}
    \centering
    \includegraphics[width=1.0\linewidth]{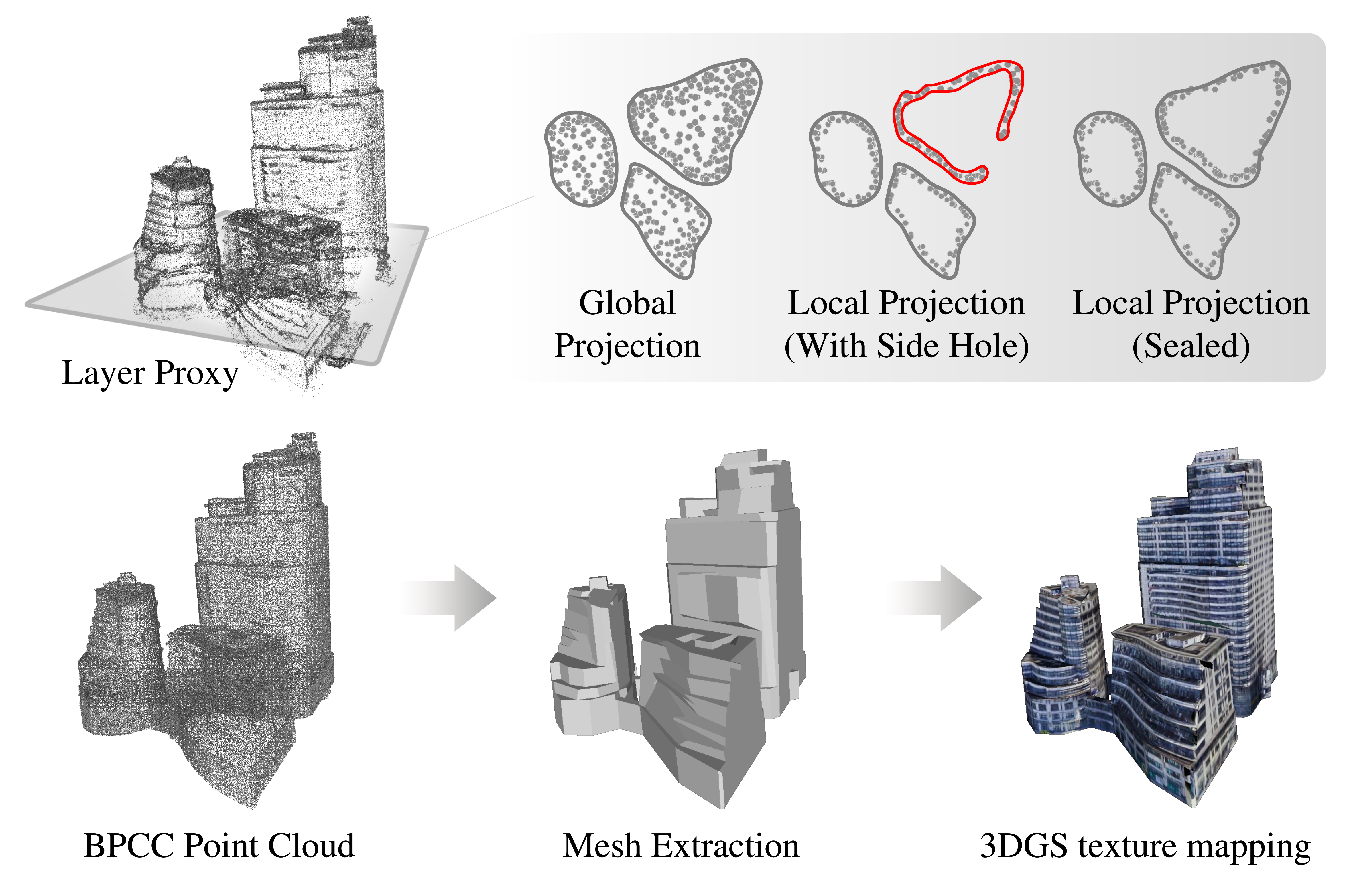}
    \caption{Textured proxy buildings Reconstruction. We first apply our proposed BPCC method to complete the bottom and side regions of the dense point cloud. We then extract the mesh and estimate textures based on renderings produced by 3DGS.}
    \label{fig:BPCC}
    \vspace{-15pt}
\end{figure}

\subsection{Mesh Extraction and Texture Mapping}

After obtaining complete and hole-free point clouds for each building, we apply the Fitting Planar Primitives (FPP) algorithm~\cite{fpp} to approximate the point clouds with planar surfaces. This method iteratively optimizes the configuration of planar primitives by minimizing the following objective function:
\begin{equation}
U(x) = w_f U_f(x) + w_s U_s(x) + w_c U_c(x),
\label{eqa_fpp}
\end{equation}
where $U_f(x), U_s(x)$ and $U_c(x)$ represent fidelity, smoothness, and completeness terms, respectively. The weights $w_f, w_s$ and $w_c$ are all set to 1, following the setting in \cite{fpp}. Subsequently, the planar primitives derived from the FPP algorithm serve as input for the Kinetic Shape Reconstruction (KSR) method~\cite{KSR}. KSR partitions space into labeled inside and outside regions, ultimately generating the proxy building geometry.

\paragraph{Texture Mapping.}
We compute color textures to enhance visual realism. Directly using aerial imagery for texture mapping often leads to visual artifacts such as seams, black patches, and misaligned textures due to color inconsistencies across views and occlusions from nearby buildings. To mitigate these problems, we generate synthetic images using 3D Gaussian models of individual buildings, providing controlled inputs that improve consistency and reduce artifacts. We then apply the method described in \cite{TwinTex} to produce high-quality textures. A key advantage of synthetic rendering is precise control over virtual camera positions and orientations, enabling accurate alignment with proxy geometries and clear depiction of architectural details such as windows, doors, and structural outlines, while minimizing occlusions. For each building, we render 28 views arranged in three vertical layers at angles from 20° to 90°. Horizontally, eight viewpoints are evenly spaced around the building at angles from $0^\circ$ to $360^\circ$, with an additional four views placed directly above to ensure comprehensive coverage. Final textures are computed at dynamic resolutions ranging from 1024 to 4096 pixels, depending on the building's bounding box size.

\paragraph{UV Finetuning.}
Although using 3D Gaussian models for texture estimation improves consistency and reduces occlusions from nearby buildings, the results can be affected by the quality of model training. To address discrepancies between the rendered proxy geometry and ground truth imagery, we refine the UV textures through differential rasterization. 

Following the standard graphics rasterization pipeline, the rendering results are sampled independently from UV maps via barycentric coordinates.
However, this one-to-one mapping deteriorates the optimization procedure for updating minor UV pixels per iteration, resulting in salt-and-pepper noise. 
We handle this problem by transforming the UV map as
\begin{equation}
    \bm{T}_\text{smooth}=\text{cov}(\bm{T}, \bm{g}),
\end{equation}
where $\bm{g}$ is the fixed convolutional kernel used to smooth textures during optimization.
As a result, the proxy building with optimized UV textures produces sharper and more realistic color renderings.

\subsection{Residual Gaussians}
Our textured proxy buildings offer a compact and efficient representation of urban structures, which occupy a significant portion of cityscapes, and enable fast rendering performance. While proxy buildings are effective for modeling simple geometry due to their reliance on polygonal representations, they struggle to capture complex appearances with high-frequency details and intricate shapes. To address this limitation, we introduce residual 3D Gaussians as a complementary component, forming a hybrid representation alongside proxy buildings. Specifically, sparse Gaussians are sampled from the initial set, guided by color residual maps (CRM) that quantify the difference between the proxy-rendered textures and the ground truth appearance, which is computed as $\bm{c}_{\text{res}} = | \bm{c}_{\text{GT}} - \bm{c}_m |$,
where $\bm{c}_m$ is the color rendered from the proxy mesh. To isolate building-specific errors, we project the proxy buildings onto the ground truth images and generate building masks to segment the structures from their backgrounds. These masks allow us to compute CRMs that highlight regions where the proxy fails to capture fine details, effectively guiding the placement of residual Gaussians for improved fidelity.

Given a triangle mesh with vertices $\bm{V}$, faces $\bm{F}$, and UV coordinates $\bm{T}$, we define the mesh as $\mathcal{M} = \{\bm{V}, \bm{F}, \bm{T}\}$. For a ray $\bm{r}_m$ passing through pixel $m$ from a specific viewpoint, the rasterization pipeline yields a color $\bm{c}_m$ sampled from the UV map and a depth value $d_m$ obtained from the z-buffer:
\begin{equation}
    d_m, \bm{c}_m=\mathcal{R}(\mathcal{M}, \bm{r}_m),
\end{equation}
where $\mathcal{R}$ denotes the rasterization operation. Since the UV map follows a surface rendering paradigm, we fix its opacity to 1.0 for compatibility with hybrid volume rendering. This allows the mesh to be treated as the final intersected surface along the ray.
Leveraging the accuracy of the proxy geometry, the mesh provides a reliable depth estimate, making $d_m$ a useful weak supervisory signal in the early stages of 3DGS training. To constrain Gaussian placement near the mesh surface, we introduce a guard interval $d_g$ , activating Gaussians with depths $d < d_m + d_g$ during hybrid rendering.
The final hybrid color $\bm{c}_h(\bm{x})$ at a point $x$ is computed by accumulating contributions from the activated Gaussians and the mesh surface:
\begin{equation}
    \bm{c}_h(\bm{x}) = \sum_{k=1}^KT_k\alpha_k\bm{c}_k + T_m\bm{c}_m,\quad T_m=\prod_{k=1}^K(1-\alpha_k).
\end{equation}
where $\alpha_k$ is the opacity and $T_k$ is the transmittance of the k-th Gaussian.
We select meaningful Gaussians based using CRM $c_{res}$ by computing a score:
\begin{equation}
    E_k^\gamma = \max_{\bm{r}\in\mathbb{P^\gamma}} c_{\text{res}}(\bm{r}) \alpha_k(\bm{r}) T_k(\bm{r}),
\end{equation}
where $\mathbb{P}^\gamma$ is the set of rays sampled corresponding to all pixels from the viewpoint-$\gamma$.
To eliminate view-dependent biases, we record the maximum score across all training views for each Gaussian:
\begin{equation}
    E_k = \max_{\gamma\in\Gamma} E_k^\gamma.
\end{equation}
This score is used to sample residual Gaussians. Gaussians with scores above a predefined threshold are selected, where the threshold can be easily adjusted based on the acceptable pixel error in the final reconstruction. 

\section{Surrounding Gaussians for Environments}
\label{sec:4}

\begin{figure*}
    \centering
    \includegraphics[width=1\linewidth]{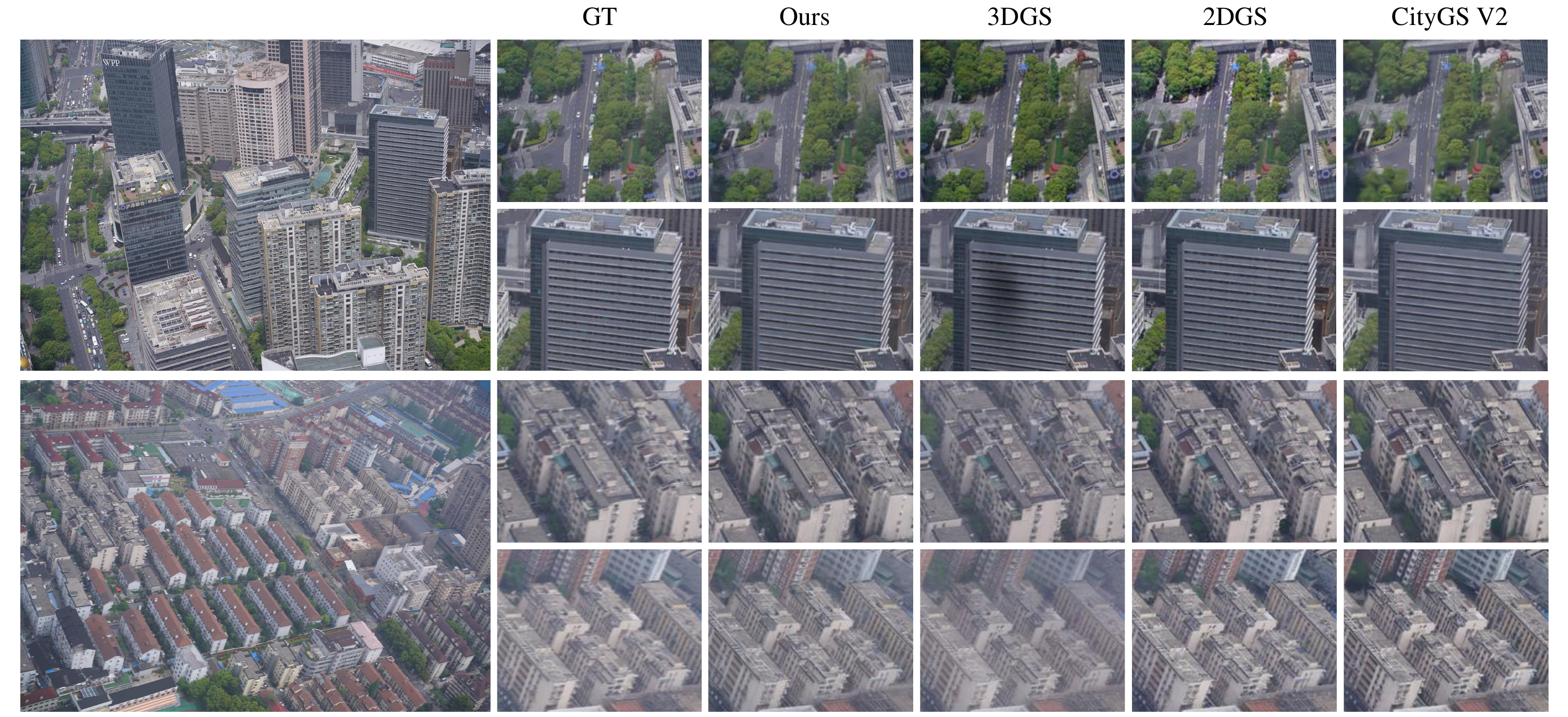}
    \caption{Qualitative comparison of our method and baseline methods on the Area-H and Area-L datasets.}
    \label{fig:Qualitative}
    \vspace{-10pt}
\end{figure*}

The surrounding environment contains many fine-grained elements, such as roads, vehicles, and trees, which are difficult to reconstruct accurately using traditional MVS methods. As a result, these elements are often oversimplified into basic proxy geometries. To better preserve their structural complexity and visual fidelity, we directly represent the surrounding environment using 3D Gaussians. Instead of using the segmented Gaussians from the initial full-scene Gaussians, we utilize a sparse subset of Gaussians downsampled from the original set to model it. To balance rendering quality with computational efficiency, we adopt the importance-based sampling strategy proposed in \cite{mini_splatting}. The importance $I_k$ of the $k$-th Gaussian is defined as the cumulative blending weight all emitted rays:
\begin{equation}
    I_k = \sum_{i=1}^N w_{k,i},
\end{equation}
where $k$ and $i$ denote the indices of Gaussians and pixels, respectively. To perform stochastic sampling that ensures an even spatial distribution of Gaussians, we assign each  Gaussian a sampling probability based on its importance:
\begin{equation}
    P_k = \frac{I_k}{\sum_{j=1}^KI_j}.
\end{equation}
Using this probability distribution, we generate a lightweight set of Gaussians to represent the environment, referred to as surrounding Gaussians. These Gaussians are then combined with the building representation, which includes both textured proxy meshes and residual Gaussians, to form a unified model of the urban scene. During the subsequent hybrid optimization stage, we refine the attributes of the surrounding Gaussians, including their position, scale, opacity, and color appearance. To preserve spatial consistency, position updates are applied using a low learning rate, ensuring minimal displacement. 
After optimization, we obtain a complete and efficient representation of the urban scene.

\section{Experimental Results}

\subsection{Training Details}
We first apply an off-the-shelf MVS method~\cite{AgisoftPhotoScan2016} to reconstruct dense point clouds from the input aerial imagery. These point clouds are then used to initialize the training of the initial 3D Gaussians, with $SH=0$ to accelerate convergence. Using the building segmentation data, we separate buildings from the surrounding environment in both the dense point clouds and in the initial Gaussian representations. To enhance the visual quality of the buildings, we refine the UV textures of the proxy meshes using the PyTorch3D framework. Residual Gaussians are extracted from the initial set using a threshold of $0.2$ to select $E_k$ and are optimized over 100K iterations. During this process, most training parameters follow those in~\cite{3DGS}, with the following exceptions: the position learning rate is reduced to $1\%$ of the original value, the densification interval is increased to $500$ steps, and densification is stopped after $25K$ iterations. For the surrounding environment, we adopt the importance-based sampling strategy to downsample the initial Gaussians by a factor of $0.1$. The resulting Gaussians are then optimized for $30,000$ iterations using the same training parameters as in~\cite{3DGS}, except that the position learning rate is again reduced to $1\%$.

\begin{table}[tb]
    \setlength{\tabcolsep}{3pt}
    \caption{Quantitative comparison on the UrbanBIS dataset.}
    \vspace{-10pt}
    \label{tab:UrbanBIS}
    \centering
    \begin{center}
        \begin{tabular*}{\linewidth}{@{\extracolsep{\fill}}p{1.6cm}ccccc@{}}
            \toprule
            {\centering Method}
            & $\text{PSNR}_\uparrow $
            & $\text{Size(MB)}_\downarrow $
            & $\text{\#GS(M)}_\downarrow $
            & $\text{Time(h)}_\downarrow $
            & $\text{FPS}_\uparrow$
            \\
            \midrule
            3DGS
            & 18.20
            & 356
            & 2.27
            & 1.6
            & 189.11

            \\
            2DGS
            & 16.85
            & 147
            & 0.93
            & 2.5
            & 12.83
            \\
            OctreeGS
            & 17.39
            & \cellcolor{color_2nd} 55
            & 1.87
            & 2.52
            & 73.10
            \\
            CityGS-v1
            & 18.38
            & 404
            & 1.71
            & 3.5
            & 130.90
            \\
            CityGS-v2
            & 17.98
            & 88
            & \cellcolor{color_2nd}0.562
            & 5.5
            & 58.65
            \\
            \midrule
            Ours
            & \cellcolor{color_2nd}18.48
            & 117
            & 0.72
            & \cellcolor{color_2nd}1.4
            & \cellcolor{color_2nd}253.86
            \\
            \bottomrule
        \end{tabular*}
    \end{center}
    \vspace{-10pt}
\end{table}


\begin{table*}[tb]
    \setlength{\tabcolsep}{3pt}
    \caption{Quantitative comparison on our Area-H and Area-L datasets.}
    \vspace{-10pt}
    \label{tab:Jingan}
    \centering
    \begin{center}
        \begin{tabular*}{\linewidth}{@{\extracolsep{\fill}}p{1.6cm}ccccc ccccc@{}}
            \toprule
            \multirow{2}{*}{\centering Method}
            & \multicolumn{5}{c}{Area-H}
            & \multicolumn{5}{c}{Area-L}
            \\
            \cmidrule[0.5pt](lr){2-6} \cmidrule[0.5pt](lr){7-11}
            {\ }
            & $\text{PSNR}_\uparrow $
            & $\text{Size~(MB)}_\downarrow $
            & $\text{\#GS~(M)}_\downarrow $
            & $\text{Time~(h)}_\downarrow $
            & $\text{FPS}_\uparrow$
            
            & $\text{PSNR}_\uparrow $
            & $\text{Size~(MB)}_\downarrow $
            & $\text{\#GS~(M)}_\downarrow $
            & $\text{Time~(h)}_\downarrow $
            & $\text{FPS}_\uparrow$
            \\
            \midrule
            3DGS
            & \cellcolor{color_2nd}22.44
            & 6319
            & 40.39
            & 24.3
            & 33.60
            & \cellcolor{color_2nd} 25.00
            & 6986
            & 44.66
            & 23.4
            & 30.93
            \\
            2DGS
            & 22.19
            & 4127
            & 27.04
            & 29.6
            & 4.59
            & 24.41
            & 3849
            & 25.22
            & 31.2
            & 5.09
            \\
            CityGS-v2
            & 21.04
            & 1140
            & 7.33
            & 30.5
            & 5.27
            & 23.95
            & 1335
            & 8.58
            & 22.8
            & 6.19
            \\
            \midrule
            Ours
            & 21.93
            & \cellcolor{color_2nd}741
            & \cellcolor{color_2nd}4.74
            & \cellcolor{color_2nd}17.6
            & \cellcolor{color_2nd}161.14
            & 23.97
            & \cellcolor{color_2nd}624
            & \cellcolor{color_2nd}3.60
            & \cellcolor{color_2nd}16.6
            & \cellcolor{color_2nd}196.00
            \\
            \bottomrule
        \end{tabular*}
    \end{center}
    \vspace{-10pt}
\end{table*}



\subsection{Comparison}

\paragraph{Datasets.}
We evaluate our framework on three real-world urban datasets: two aerial datasets captured using drones and the publicly available UrbanBIS~\cite{UrbanBIS} dataset. Our aerial datasets consist of two distinct scenes, each covering an area of $1.5$ km² and characterized by different architectural typologies. One scene, referred to as Area-H, primarily features high-rise buildings in dense commercial zones, while the other, Area-L, is dominated by low-rise structures typical of residential areas. 
Both datasets present considerable challenges due to complex geometries such as rooftops and facades, as well as diverse material appearances including glass, concrete, and metal. 
Area-H consists of 8,047 images, while Area-L contains 6,192 images, with 87.5\% allocated to the training set and the remaining images reserved for testing.



\paragraph{Competing Methods.}

\begin{table}[h]
\centering
\caption{Rendering Performance (FPS) on NVIDIA Jetson AGX Orin at 720p resolution.}
\vspace{-10pt}
\begin{tabular*}{\linewidth}{@{\extracolsep{\fill}}cccc@{}}
\toprule
 & Area-H & Area-L & UrbanBIS \\ \midrule
3DGS & 5 & 6 & 29 \\ 
Ours & \cellcolor{color_2nd}20 & \cellcolor{color_2nd}24 & \cellcolor{color_2nd}51 \\ 
\bottomrule
\end{tabular*}
\label{tab:jetson}
\vspace{-10pt}
\end{table}

We compare our CityGo framework with 3DGS~\cite{3DGS}, 2DGS~\cite{2DGS}, OctreeGS~\cite{OctGS}, CityGS-V1~\cite{CityGS} and CityGS-V2~\cite{CityGS_v2}. 
For CityGo, 3DGS and 2DGS, we adopt a divide-and-conquer strategy by partitioning large urban scenes into overlapping blocks with a $20\%$ overlap. Each block is trained independently for $100K$ iterations, using $SH=2$ for 3DGS and 2DGS, and $SH=0$ for our initial Gaussians. After training, overlapping regions are cropped to their bounding boxes and seamlessly merged to reconstruct the complete scene. For OctreeGS, CityGS-V1, and CityGS-V2, we follow their default partitioning strategies. All experiments are conducted on an NVIDIA RTX A6000 GPU.

\begin{table}[tb]
    \vspace{5pt}
    \centering
    \caption{Ablation results on UrbanBIS.}
    \vspace{-10pt}
        \begin{tabular*}{\linewidth}{@{\extracolsep{\fill}}p{2.1cm}cccc@{}}
            \toprule
            {\centering w/o}
            & $\text{PSNR}_\uparrow $
            & $\text{Size}_\downarrow $
            & $\text{\#GS(M)}_\downarrow $
            & $\text{FPS}_\uparrow$
            \\
            \midrule
            UV Finetuning
            & 18.45
            & 126.05
            & 0.81
            & 244.05 
            \\
            CRM
            & 18.33
            & \cellcolor{color_2nd}90.23
            & \cellcolor{color_2nd}0.58
            & 238.58 
            %
            \\ 
            \midrule
            Ours
            & \cellcolor{color_2nd}18.48
            & 112.27
            & 0.72
            & \cellcolor{color_2nd}256.38 
            \\
            \bottomrule
        \end{tabular*}
    \label{tab:ablation}
    \vspace{-10pt}
\end{table}

\paragraph{Qualitative Comparison.}
As show in Fig.~\ref{fig:Qualitative}, our method reduces the model size to approximately 1/8 of 3DGS while preserving most of the visual details. 3DGS tends to generate floating Gaussian points around scene surfaces. Additionally, due to the training strategy of dividing the scene into blocks and then merging them, which is adopted by methods like 3DGS and 2DGS, color discrepancies between blocks are inevitable, as can be clearly observed in the Fig.\ref{fig:Qualitative}. Our method mitigates the floating Gaussian issue present in 3DGS and eliminates the color discrepancies between blocks caused by block-based training in large scenes. By removing these visually distracting artifacts, our method provides a superior visual experience.

\paragraph{Quantitative Comparison.} 

\begin{table}[tb]
    \vspace{5pt}
    \centering
    \caption{Summary of Area-H Building Meshes.}
    \vspace{-10pt}
    \begin{tabular*}
    {\linewidth}{@{\extracolsep{\fill}}ccccc@{}}
        \toprule
        Area-H & Size (MB) & Vertices & Faces \\ \midrule
        MVS Meshes  & 3329.24 & 14232977 & 28462751 \\
        Proxy Meshes  & \cellcolor{color_2nd}53.96 & \cellcolor{color_2nd}33106 & \cellcolor{color_2nd}57374 \\ \bottomrule
    \end{tabular*}
    \label{tab:mesh_data}
    \vspace{-15pt}
\end{table}

For quantitative comparison, we evaluate the visual metric PSNR, storage comsumption via model size, memory comsumption via Gaussians counts, training time and rendering speed (FPS at $1988\times1326$ resolution).
Table~\ref{tab:UrbanBIS} presents the comparison results on the UrbanBIS dataset, demonstrating that our method outperforms others in both rendering quality and speed.
We further evaluate our method on the Area-H and Area-L datasets, as shown in Table~\ref{tab:Jingan}. CityGo achieves the highest rendering speed, reaching $161.14$ FPS on Area-H and $196$ FPS on Area-L. 
It also yields the smallest model size and the shortest training time among all compared methods. While there is a slight reduction in PSNR, with a maximum drop of $1.03$ dB compared to 3DGS, the significant improvements in efficiency make CityGo a highly competitive solution. 
To further evaluate the efficiency of our hybrid representation, we also compare CityGo and 3DGS on the mobile NVIDIA Jetson AGX Orin GPU at 720p resolution (a typical screen resolution for intelligent vehicles), with results summarized in Table~\ref{tab:jetson}. CityGo consistently outperforms in rendering performance across all three datasets, enabling real-time rendering of large-scale urban scenes up to 1.5 km² at a minimum of 20 FPS.


\paragraph{Proxy Mesh.}
Our proxy buildings provide a highly lightweight and compact representation of the underlying building geometries. Table.~\ref{tab:mesh_data} presents a comparison with MVS-generated meshes in terms of file size, number of vertices, and number of faces. The proxy building achieves a compression factor of $\times 61.7$ in mesh size when compared to the MVS-generated mesh.

\begin{figure}
    \centering
    \includegraphics[width=1\linewidth]{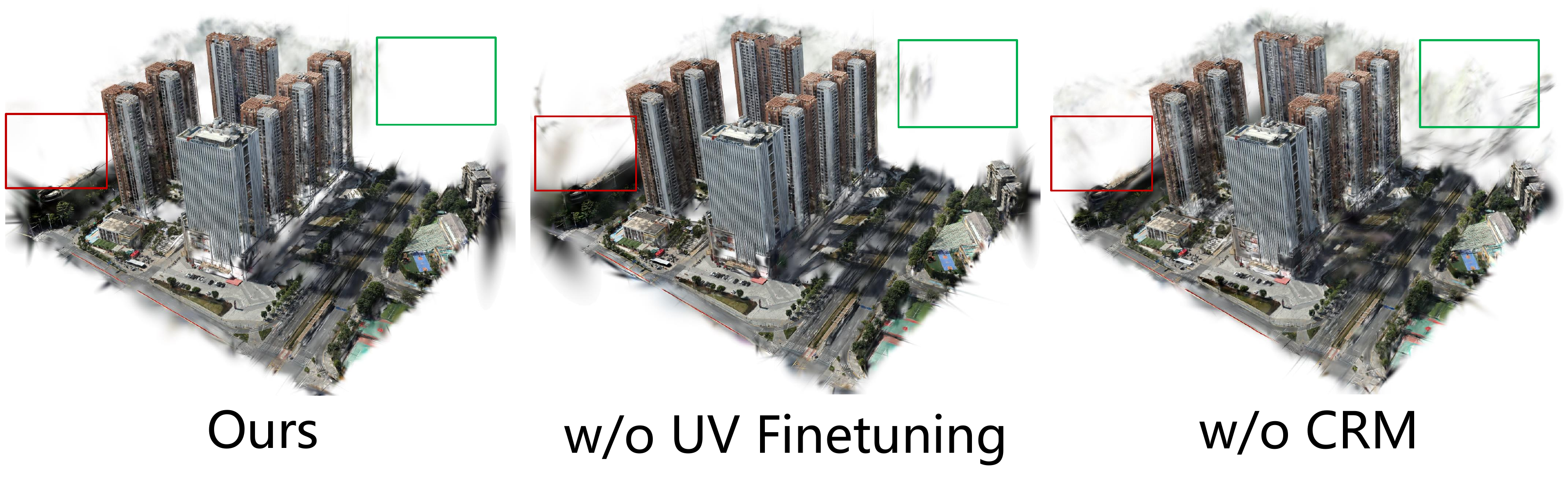}
    \caption{Ablation studies on UV finetuning and CRM-based sampling demonstrate that the proposed UV finetuning and CRM methods result in superior visual quality, achieving cleaner output compared to competing approaches.}
    \label{fig:ablation}
    \vspace{-20pt}
\end{figure}


\subsection{Ablations}

\paragraph{UV Finetuning and CRM based Sampling.} Table~\ref{tab:ablation} presents the quantitative ablation results for our UV finetuning scheme and the CRM-based sampling strategy, while the corresponding qualitative results are shown in Fig.~\ref{fig:ablation}.
Without UV finetuning, more residual Gaussians are required to compensate for the appearance discrepancies, resulting in increased model size. In the absence of CRM-based sampling, replacing it with random sampling degrades rendering quality and introduces noticeable Gaussian floaters



\section{Conclusion and Limitations}
This paper presents a lightweight hybrid reconstruction framework designed for large-scale urban scenes, balancing visual quality with rendering efficiency. Our method combines textured proxy meshes and residual Gaussians for buildings, while using compact 3D Gaussian Splatting (3DGS) for distant and unstructured elements. A carefully crafted training strategy ensures high fidelity and significantly reduces model size compared to existing 3DGS-only approaches.

\begin{figure}
    \centering
    \includegraphics[width=1.0\linewidth]{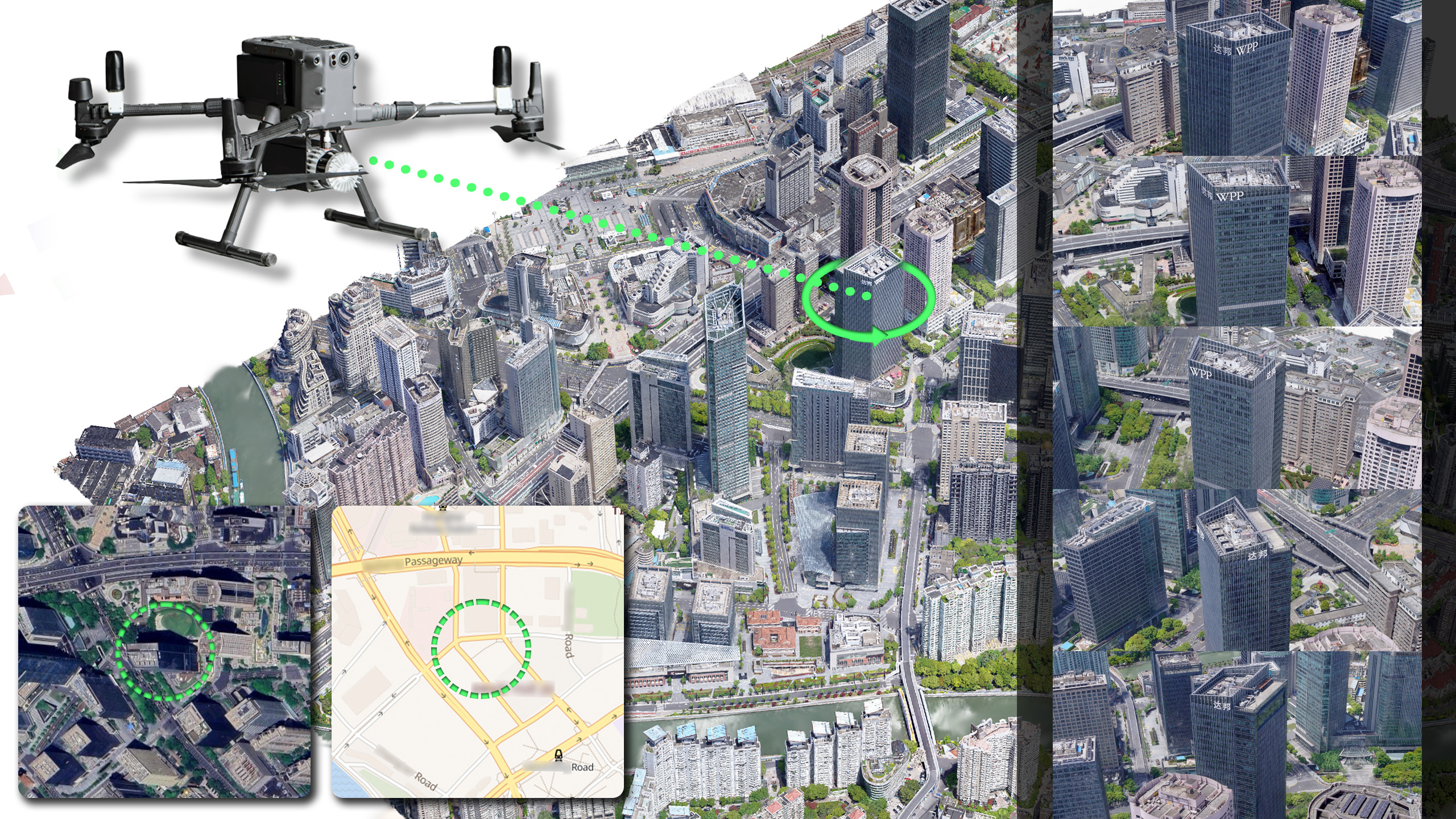}
    \caption{Practical Use Cases of CityGo in Urban Planning.}
    \label{fig:app}
\end{figure}

The framework enables real-time, cinematic rendering on resource-limited devices, where bandwidth and computational resources are typically constrained. As shown in Fig.~\ref{fig:app}, this capability is crucial for applications in urban planning, autonomous navigation, and aerial delivery, advancing neural rendering toward real-world use. Our work provides a practical solution for scalable, photorealistic urban digital twins.

However, the system's reliance on accurate proxy geometry poses challenges. Non-building structures, like cranes or signage, may be misclassified as buildings, and the fixed-opacity textures (set to $1.0$) prevent 3DGS from correcting such errors, resulting in artifacts and PSNR degradation. Future work will explore semantic-aware modeling and adaptive transparency to address these issues.

{
    \small
    \bibliographystyle{ieeenat_fullname}
    \bibliography{main}
}

\appendix
\section{Building Point Cloud Completion Algorithm}
\label{app:bccp}

Here is the detailed algorithm of BPCC. Note that  for the convenience of expression, unlike in the main text, the first layer is the bottom layer and the $L$th layer is the top layer. $\mathbf{Conc}$ means extracting the outer contour of the alpha-shape from a point set. $\mathbf{Clust}$ means clustering a point set.
\begin{algorithm} 
\KwData{Dense point cloud $\mathcal{P}$ of a building from MVS and the number of Layers $L$}
\KwResult{Building point cloud $\mathcal{C}$ with downside and holes closed}
Initialization the point cloud $\mathcal{C} \gets \mathcal{P}$\;
Initialization the global projected point set $\cup_{i=1}^{L}P_i$\;
Initialization the local projected point set $\cup_{i=1}^{L}D_i$\;
Initialization the dominant structural profile set $\mathcal{S} \gets \emptyset$\;
$\mathcal{P}_1 \gets \mathbf{Clust}(P_1)$\;
\For{$P_1^k \in \mathcal{P}_1$}{
    $C_1^k \gets \mathbf{Conc}(P_1^k)$\;
    $\mathcal{S} \gets \mathcal{S} \cup \{C_1^k\}$\;
    $\hat{S}_1^k = C_1^k$
}

\caption{Building Point Cloud Completion}
\label{algo_proxy_pcd}
\end{algorithm}

\begin{algorithm}[!t]
\KwData{Dense point cloud $\mathcal{P}$ of a building from MVS and the number of Layers $L$}
\KwResult{Building point cloud $\mathcal{C}$ with downside and holes closed}
\For{$i\leftarrow 2$ \KwTo $L$}{
    $\mathcal{P}_i \gets \emptyset$\;
    \For{$P_{i-1}^k \in \mathcal{P}_{i-1}$}{
        $\hat{P}_i^k \gets P_{i-1}^k \cap P_{i}$\;
        $\mathcal{P}_i^k \gets \mathbf{Clust}(\hat{P}_i^k)$\;
        \eIf{$\|\mathcal{P}_i^k\| = 1$}{
            $C_i^m \gets \mathbf{Conc}(\mathbf{P}_{i}')$\;
            \eIf{$\mathbf{Area}(C_i^m) / \mathbf{Area}(\hat{S}_{i-1}^k) \leq \gamma$}{
                $\mathcal{S} \gets \mathcal{S} \cup \{C_i^m\}$\;
                $\hat{S}_i^m \gets C_i^m$\;
                $\mathbf{Lmax}(\hat{S}_{i-1}^k) \gets i$\;
                $m \gets m + 1$\;
            }
            {
                $\hat{S}_i^m \gets \hat{S}_{i-1}^k$\;
            }
        }{
            \For{$\mathbf{P}_i^k \in \mathcal{P}_i^k$}{
                $C_i^m \gets \mathbf{Conc}(P_i^k)$\;
                $\mathcal{S} \gets \mathcal{S} \cup \{C_i^m\}$\;
                $\hat{S}_i^m \gets C_i^m$\;
                $\mathbf{Lmax}(\hat{S}_{i-1}^k) \gets i$\;
                $m \gets m + 1$\;
            }
        }
    }
}
\For{$S_i \in \mathcal{S}$}{
    Extrude polyhedral cell from $S_i$ to the layer $\mathbf{Lmax}(S_{i-1}^k)$\;
}
Stack all polyhedral cells to form proxy geometry\;
Sample points at layer $1$ of proxy geometry downside and append to $\mathcal{C}$\;
\For{$i\leftarrow 1$ \KwTo $L$}{
    \For{$P_i^k \in \mathcal{P}_i$}{
        $D_i^k \gets P_i^k \cap D_{i}$\;
        $F_i^k \gets \mathbf{Conc}(D_i^k)$\;
        \If{$\mathbf{Area}(F_i^k) / \mathbf{Area}(C_i^k) < \beta$}{
            Sample points at layer $i$ of polyhedral cell surface of $\hat{S}_i^k$ and append to $\mathcal{C}$\;
        }
    }
}
\caption{Building Point Cloud Completion (Continued)}
\label{algo:proxy_pcd2}
\end{algorithm}

\begin{figure*}
    \centering
    \includegraphics[width=1\linewidth]{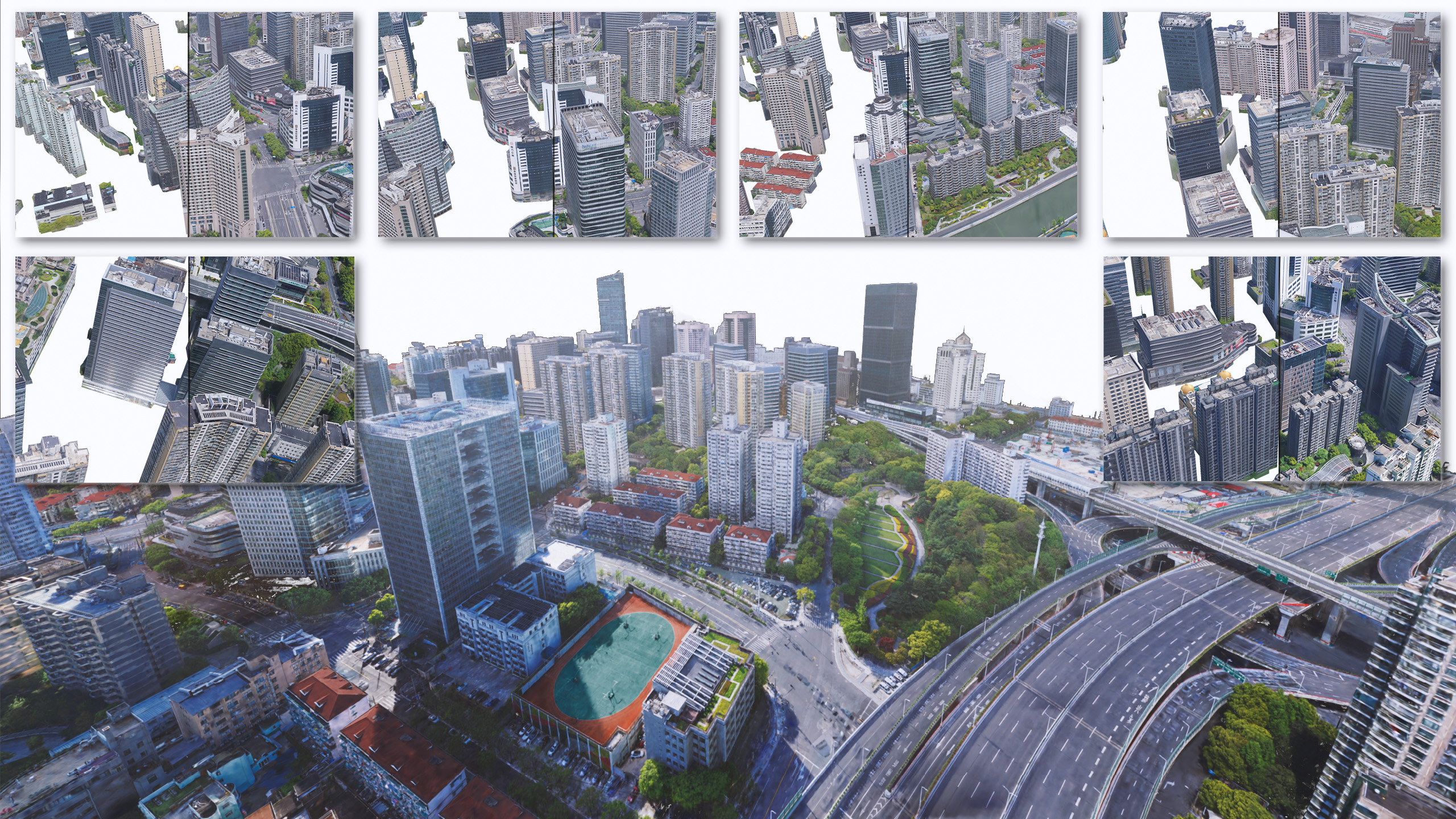}
    \caption{Visualization of examples rendered using our CityGO models in the Area-H scene. Comparisons are provided to highlight the proxy building and the final rendered results.
    }
    \label{fig:Gallery1}
\end{figure*}
\begin{figure*}
    \centering
    \includegraphics[width=1\linewidth]{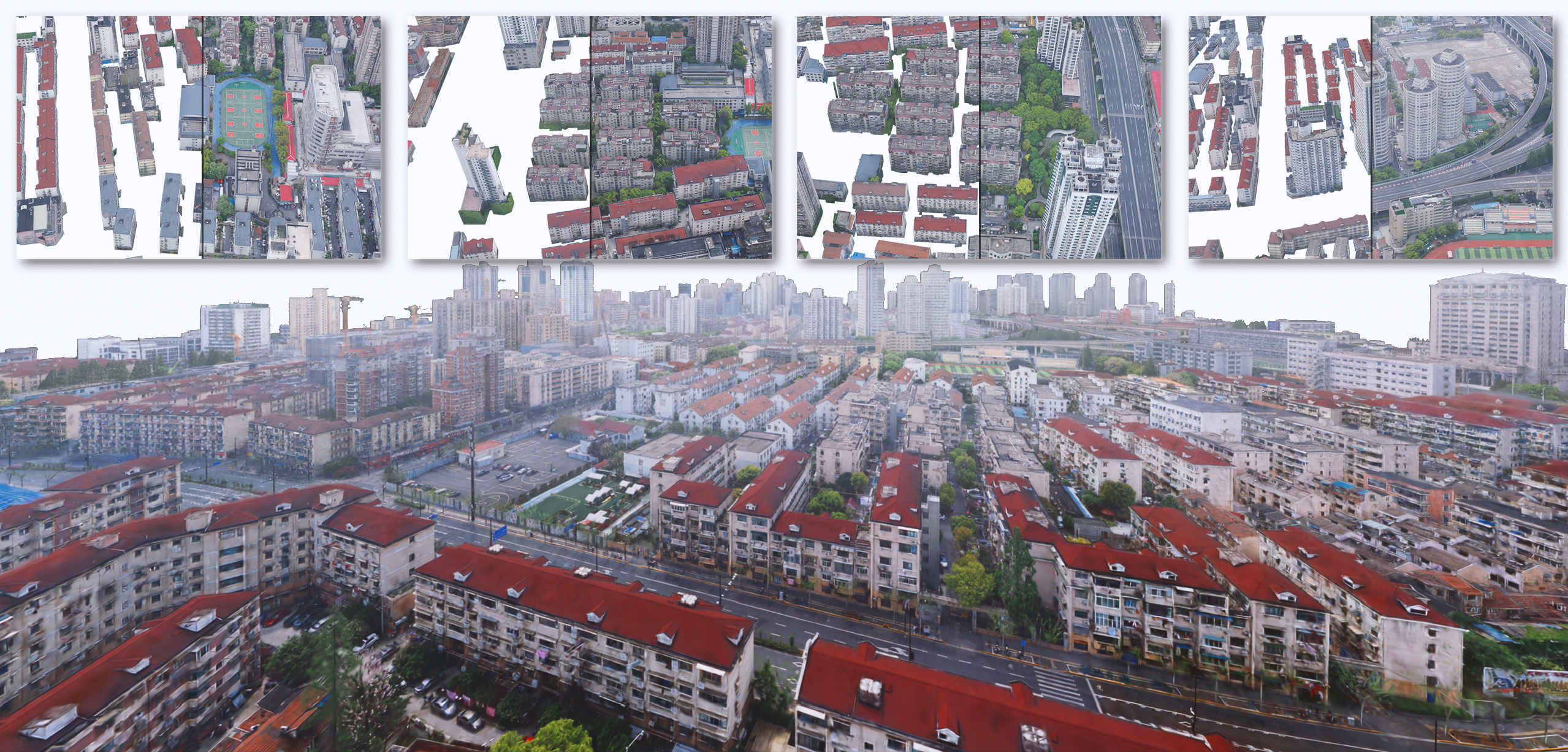}
    \caption{Visualization of examples rendered using our CityGO models in the Area-L scene. Comparisons are provided to highlight the proxy building and the final rendered results.}
    \label{fig:Gallery2}
\end{figure*}

\end{document}